\documentclass[12pt]{article}
\usepackage{graphicx}
\usepackage{epstopdf}
\usepackage{amsmath,amssymb,amsfonts,graphicx}

\begin{document}
\thispagestyle{empty}
\renewcommand{\refname}{References}

\title{\bf Hot dense magnetized ultrarelativistic spinor matter in a slab} 

\author{Yurii A. Sitenko}

\date{}

\maketitle
\begin{center}
Bogolyubov Institute for Theoretical Physics,\\
National Academy of Sciences of Ukraine,\\
14-b Metrologichna Street, 03680 Kyiv, Ukraine

\end{center}

\begin{abstract}
Properties of hot dense ultrarelativistic spinor matter in a slab of finite width, placed in a transverse uniform magnetic field, 
are studied. The admissible set of boundary conditions is determined by the requirement that spinor matter be confined inside the 
slab. In thermal equilibrium, the chiral separation effect in the slab is shown to depend on both temperature and chemical 
potential; this is distinct from the unrealistic case of the magnetic field filling the unbounded (infinite) medium, when the 
effect is temperature independent. In the realistic case of the slab, a stepwise behaviour of the axial current density at zero temperature is 
smoothed out as temperature increases, turning into a linear behaviour at infinitely large temperature. A choice of boundary 
conditions can facilitate either augmentation or attenuation of the chiral separation effect; in particular, the effect can persist even at zero 
chemical potential, if temperature is finite. Thus the boundary condition can serve as a source that is additional to the spinor 
matter density. 
\end{abstract}


\bigskip

\begin{center}
\end{center}

\bigskip

\section{Introduction}

The effects of background fields in quantum field theory are extensively analyzed from various perspectives. Recent studies of the 
influence of a strong background magnetic field on properties of hot dense relativistic quantized matter have drawn the attention 
of researchers in diverse areas of contemporary physics, ranging from cosmology, astroparticle and high energy physics to 
condensed matter physics. Relativistic heavy-ion collisions \cite{Khar}, compact astrophysical objects (neutron stars and 
magnetars) \cite{Char}, the early universe \cite{Tash}, novel materials known as the Dirac and Weyl semimetals \cite{Liu,Vaf} 
are the main physical systems where these studies are relevant. A source of the background magnetic field can be different, 
varying from one system to another. For condensed 
matter systems, one may simply apply an external magnetic field either to probe their physical properties or to better 
understand the underlying physics. In other cases, a generation of the magnetic field is the inherent property of the system. In 
the case of relativistic heavy-ion collisions, very strong magnetic fields are produced during the early stages of the collision 
as a result of electric currents from the colliding charged ions \cite{Sko}; because of the high electric conductivity of the 
medium, the appropriate fields may survive for as long as the lifetime of the quark-gluon plasma itself and, thus, have a profound 
impact on the plasma dynamics. In the case of compact stars, the existence of strong magnetic fields is inferred from observational 
data \cite{Ola}; even though the exact nature of the underlying mechanism responsible for a generation of such fields may still be 
debated, it is undoubted that such fields exist and play a significant role in the stellar physics. In the case of the early 
universe, several competing mechanisms were proposed, providing for a generation of very strong magnetic fields \cite{Gra}; despite 
the difference in details, the consensus is that rather strong magnetic fields should have been generated, since this is required 
by the present-day observation of weak, but nonvanishing, intergalactic magnetic fields. 

Assuming that temperature and chemical potential, as well as the inverse magnetic length, exceed considerably the mass of a 
relativistic quantized spinor matter field, a variety of chiral effects emerges in hot dense magnetized matter in thermal 
equilibrium; the lowest Landau level is primarily responsible for this; see review in \cite{Mir} and references 
therein. One of the basic effects is the chiral separation effect that is characterized by the nondissipative axial 
current along the direction of the magnetic field strength, ${\bf B}$, \cite{Vil,Son,Met} 
\begin{equation}
{\bf J}^5=-\frac{e{\bf B}}{2\pi^2}\mu; \label{eq1}
\end{equation}
note that the current is linear in chemical potential $\mu$ and spinor particle charge $e$, being independet of temperature $T$. 

So far chiral effects were mostly considered in unbounded (infinite) matter, which may be relevant for cosmological applications, 
perhaps. For all other applications (to astroparticle, high energy and condensed matter physics), an account has to be taken of the 
finiteness of physical systems, and the role of boundaries in chiral effects in bounded matter has to be clearly exposed. The 
concept of quantized matter fields which are confined to bounded spatial regions is quite familiar in the context of condensed 
matter physics: collective excitations (e.g., spin waves and phonons) exist only inside material samples and do not spread outside. 
Nevertheless, a quest for boundary conditions ensuring the confinement of quantized matter was initiated in particle physics, in 
the context of a model description of hadrons as bags containing quarks \cite{Bog, Cho1}. Motivations for a concrete form of the 
boundary condition may differ in detail, but the key point is that the boundary condition has to forbid any 
flow of quark matter across the boundary, see \cite{Joh}. However, from this point of view, the bag boundary conditions 
proposed in \cite{Bog, Cho1} are not the most general ones. It has been rather recently realized that the most general boundary 
condition ensuring the confinenent of relativistic quantized spinor matter within a simply connected boundary involves four 
arbitrary parameters \cite{Bee, Wie}, and the explicit form of such a condition has been given \cite{Si1, Si2, Si3}. To study an 
impact of the background magnetic field on confined matter, one has to choose the magnetic field configuration with respect to the 
boundary surface. The primary interest is to understand the effect of a boundary which is transverse to the magnetic field strength 
lines. Then the simplest geometry is that of a slab in the uniform magnetic field directed perpendicular. It should 
be noted that such a geometry can be realized in condensed matter physics by putting slices of Dirac or Weyl semimetals in an 
external transverse magnetic field. Note also that the slab geometry is conventional in a setup for the Casimir effect 
\cite{Cas1}, see review in \cite{Bor}. 

As a first step toward the full theory of chiral effects in bounded matter, the authors of \cite{Gor} considered the chiral 
effects in dense magnetized ultrarelativistic spinor matter at zero temperature in a slab with the use of the bag 
boundary condition of \cite{Bog}. The aim of the present paper is to extend the consideration to the case of nonzero temperature 
and the most general boundary condition.

In the next section, some basics of the formalism of quantum field theory in thermal equilibrium are reviewed. A choice of the 
boundary condition is discussed in Section 3. The chiral effects in the slab geometry are considered in Section 4. The conclusions 
are drawn and discussed in Section 5. The solution to the Dirac equation in the background uniform magnetic field in the slab 
geometry is given in Appendix A. In Appendix B, we calculate some sums over values of the wave number vector in the direction 
along the magnetic field.

\section{Preliminaries. Quantized spinor matter in thermal equilibrium}

We start with the operator of the second-quantized spinor field in a static background,
\begin{equation}
\hat{\Psi}({\bf r},t)=\sum\limits_{E_\lambda>0}e^{-{\rm i}E_\lambda t}\left\langle {\bf r}|\lambda\right\rangle\hat{a}_\lambda+\sum\limits_{E_\lambda<0}e^{-{\rm i}E_\lambda t}\left\langle {\bf r}|\lambda\right\rangle\hat{b}_\lambda^{\dagger},\label{eq2}
\end{equation}
where $\hat{a}_\lambda^{\dagger}$ and $\hat{a}_\lambda$ ($\hat{b}_\lambda^{\dagger}$ and $\hat{b}_\lambda$) are the spinor particle (antiparticle) creation and destruction operators satisfying  anticommutation relations,
\begin{equation}
\left[\hat{a}_\lambda,\,\hat{a}_{\lambda'}^{\dagger}\right]_+ = \left[\hat{b}_\lambda,\,\hat{b}_{\lambda'}^{\dagger}\right]_+=\left\langle \lambda|\lambda'\right\rangle, \label{eq3}
\end{equation}
and $\left\langle {\bf r}|\lambda\right\rangle$ is the solution to the stationary Dirac equation,
\begin{equation}
H\left\langle {\bf r}|\lambda\right\rangle=E_\lambda\left\langle {\bf r}|\lambda\right\rangle,\label{eq4}
\end{equation}
$H$ is the Dirac Hamiltonian, $\lambda$ is the set of parameters (quantum numbers) specifying a one-particle state, $E_\lambda$ is the energy of the state; wave functions $\left\langle {\bf r}|\lambda\right\rangle$ satisfy the requirement of orthonormality 
\begin{equation}
\int\limits_{\Omega}{\rm d}^3r\left\langle \lambda|{\bf r}\right\rangle\left\langle {\bf r}|\lambda'\right\rangle=\left\langle \lambda|\lambda'\right\rangle\label{eq5}
\end{equation}
and completeness
\begin{equation}
\sum\left\langle {\bf r}|\lambda\right\rangle\left\langle \lambda|{\bf r}'\right\rangle=I\delta({\bf r}-{\bf r}');\label{eq6}
\end{equation}
summation is over the whole set of states, and $\Omega$ is the quantization volume. 

Conventionally, the operators of dynamical variables (physical observables) in second-quantized theory are defined as bilinears of the fermion field operator (2). One can define the fermion number operator,
\begin{equation}
\hat{N}=\frac{1}{2}\int\limits_{\Omega}{\rm d}^3r(\hat{\Psi}^{\dagger}\hat{\Psi}-\hat{\Psi}^T\hat{\Psi}^{\dagger T})=
\sum\left[\hat{a}_\lambda^{\dagger}\hat{a}_\lambda-\hat{b}_\lambda^{\dagger}\hat{b}_\lambda-\frac{1}{2}{\rm sgn}(E_\lambda)\right], \label{eq7}
\end{equation}
and the energy (temporal component of the energy-momentum vector) operator,
\begin{equation}
\hat{P}^0=\frac{1}{2}\int\limits_{\Omega}{\rm d}^3r(\hat{\Psi}^{\dagger}H\hat{\Psi}-\hat{\Psi}^TH^T\hat{\Psi}^{\dagger T})=
\sum|E_\lambda|\left(\hat{a}_\lambda^{\dagger}\hat{a}_\lambda+\hat{b}_\lambda^{\dagger}\hat{b}_\lambda-\frac{1}{2}\right),\label{eq8}
\end{equation}
where superscript $T$ denotes a transposition and ${\rm sgn}(u)$ is the sign function [${\rm sgn}(\pm u)=\pm 1$ at $u>0$]. Let us introduce partition function
\begin{equation}
Z(T,\mu)={\rm Sp} \, {\rm exp}\left[-(\hat{P}^0-\mu\hat{N})/T\right], \label{eq9}
\end{equation}
where equilibrium temperature $T$ is defined in the units of the Boltzmann constant, and ${\rm Sp}$ denotes the trace or the sum 
over the expectation values in the Fock state basis created by operators in (3). Then the average of operator $\hat{U}$ over the 
grand canonical ensemble is defined as (see, e.g., \cite{Das})
\begin{equation}
\left\langle \hat{U}\right\rangle_{T,\mu}=Z^{-1}(T,\mu) \, {\rm Sp} \, \hat{U}{\rm exp}\left[-(\hat{P}^0-\mu\hat{N})/T\right]. \label{eq10}
\end{equation}
In particular, one can compute averages 
\begin{equation}
\left\langle \hat{a}_\lambda^{\dagger}\hat{a}_\lambda\right\rangle_{T,\mu}=\left\{{\rm exp}[(E_\lambda-\mu)/T]+1\right\}^{-1}, \quad E_\lambda>0 \label{eq11}
\end{equation}
and
\begin{equation}
\left\langle \hat{b}_\lambda^{\dagger}\hat{b}_\lambda\right\rangle_{T,\mu}=\left\{{\rm exp}[(-E_\lambda+\mu)/T]+1\right\}^{-1}, \quad E_\lambda<0. \label{eq12}
\end{equation}

Let us consider an operator in the form
\begin{equation}
\hat{U}=\frac{1}{2}\left(\hat{\Psi}^{\dagger}\Upsilon\hat{\Psi}-\hat{\Psi}^T\Upsilon^T\hat{\Psi}^{\dagger T}\right),\label{eqno13}
\end{equation}
where $\Upsilon$ is an element of the Dirac-Clifford algebra. The explicit form of $\hat{U}$, $\hat{P}^0$ and $\hat{N}$ in terms 
of the creation and destruction operators is inserted in (10); then, using (11) and (12), one obtains
\begin{equation}
\left\langle \hat{U}\right\rangle_{T,\mu}=-\frac{1}{2}{\rm tr}\left\langle {\bf r}|\Upsilon\tanh [(H-\mu I)(2T)^{-1}]|{\bf r}\right\rangle,\label{eq14}
\end{equation}
where ${\rm tr}$ denotes the trace over spinor indices. One can define the vector current density, 
\begin{equation}
{\bf J}=\left\langle \hat{U}\right\rangle_{T,\mu}\biggr|_{\Upsilon=\gamma^0\boldsymbol{\gamma}},\label{eq15}
\end{equation}
the axial current density,
\begin{equation}
{\bf J}^5=\left\langle \hat{U}\right\rangle_{T,\mu}\biggr|_{\Upsilon=\gamma^0\boldsymbol{\gamma}\gamma^5},\label{eq16}
\end{equation}
and the axial charge density,
\begin{equation}
J^{05}=\left\langle \hat{U}\right\rangle_{T,\mu}\biggr|_{\Upsilon=\gamma^5},\label{eq17}
\end{equation}
where $\gamma^5=-{\rm i}\gamma^0\gamma^1\gamma^2\gamma^3$ ($\gamma^0$, $\gamma^1$, $\gamma^2$, and $\gamma^3$ are the generating elements of the Dirac-Clifford algebra, and $\gamma^5$ is defined according to \cite{Okun}).

To study an influence of a background magnetic field on the properties of hot dense spinor matter, one has to account for the 
fact that the realistic physical systems are bounded. Our interest is in an effect of the static magnetic field with strength lines 
which are orthogonal to a boundary. Then the simplest geometry of a material sample is that of a straight slab in the uniform 
magnetic field directed perpendicular. Assuming that the magnetic field is strong (supercritical), we are considering 
ultrarelativistic spinor matter at high temperature and high density; thus the mass of the spinor matter field is neglected. The 
Dirac Hamiltonian takes form 
\begin{equation}
H=-{\rm i}\gamma^0\boldsymbol{\gamma}\cdot(\boldsymbol{\partial}-{\rm i}e{\bf A}),\label{eq18}
\end{equation}
and the one-particle energy spectrum is 
\begin{equation}
E_{nl}=\pm\omega_{nl},\,\,\,\,\omega_{nl}=\sqrt{2n|eB|+k_l^2},\,\,\,\,n=0,1,2,\ldots,\label{eq19}
\end{equation} 
where $B$ is the value of the magnetic field strength, ${\bf B}=\boldsymbol{\partial}\times{\bf A}$, $n$ labels the Landau levels, 
and $k_l$ is the value of the wave number vector along the magnetic field; the set of the $k_l$ values is to be determined by the 
boundary condition.

\section{Choice of boundary conditions}

The most general boundary condition ensuring the confinement of relativistic spinor matter within a simply connected boundary 
is (see \cite{Si2,Si3})
\begin{equation}
\left\{I-\gamma^0\left[{\rm e}^{{\rm i}\varphi\gamma^5}\cos\theta + (\gamma^1\cos\varsigma + 
\gamma^2\sin\varsigma)\sin\theta\right]{\rm e}^{{\rm i}\tilde{\varphi}\gamma^0(\boldsymbol{\gamma}\cdot\boldsymbol{n})}\right\}
\chi(\mathbf{r})\left.
\right|_{\mathbf{r}\in \partial\Omega}=0,\label{eq20}
\end{equation}
where $\boldsymbol{n}$ is the unit normal to surface $\partial\Omega$ bounding spatial region $\Omega$ 
and $\chi(\mathbf{r})$ is the confined spinor matter wave function, $\mathbf{r} \in \Omega$; matrices $\gamma^1$ and $\gamma^2$ in (20) are 
chosen to obey condition
\begin{equation}
[\gamma^1,\,\boldsymbol{\gamma}\cdot\boldsymbol{n}]_+=
[\gamma^2,\,\boldsymbol{\gamma}\cdot\boldsymbol{n}]_+=[\gamma^1,\,\gamma^2]_+=0,\label{eq21}
\end{equation}
and the boundary parameters in (20) are chosen to vary as
\begin{equation}
-\frac{\pi}{2}<\varphi\leq\frac{\pi}{2}, \quad -\frac{\pi}{2}\leq\tilde{\varphi}<\frac{\pi}{2}, \quad 
0\leq\theta<\pi, \quad 0\leq\varsigma<2\pi. \label{eq22}
\end{equation}
The Massachusetts Institute of Technology (MIT) bag boundary condition \cite{Joh},
\begin{equation}
(I+{\rm i}\boldsymbol{\gamma}\cdot\boldsymbol{n})\chi(\mathbf{r})\left.\right|_{\mathbf{r}\in \partial\Omega}=0, \label{eq23}
\end{equation}
is obtained from (20) at $\varphi=\theta=0$, $\tilde{\varphi}=-{\pi}/{2}$.

The boundary parameters in (20) can be interpreted as the self-adjoint extension parameters. The self-adjointness of the 
one-particle energy (Dirac Hamiltonian in the case of relativistic spinor matter) operator in first-quantized theory is required 
by general principles of comprehensibility and mathematical consistency; see \cite{Neu}. To put it simply, a multiple action is 
well defined for a self-adjoint operator only, allowing for the construction of functions of the operator, such as resolvent, 
evolution, heat kernel and zeta-function operators, with further implications upon second quantization.

In the case of a disconnected boundary consisting of two simply connected components, 
$\partial\Omega=\partial\Omega^{(+)}\bigcup\partial\Omega^{(-)}$, there are in general eight boundary parameters: 
$\varphi_{+}$, $\tilde{\varphi}_{+}$, $\theta_+$, and $\varsigma_+$ corresponding to $\partial\Omega^{(+)}$; and 
$\varphi_{-}$, $\tilde{\varphi}_{-}$, $\theta_{-}$, and $\varsigma_-$ corresponding to $\partial\Omega^{(-)}$. If spatial region $\Omega$ has the 
form of a slab bounded by parallel planes, $\partial\Omega^{(+)}$ and $\partial\Omega^{(-)}$, separated by distance $a$, then the boundary condition takes form  
\begin{equation}
\left(I-K^{(\pm)}\right)
\chi(\mathbf{r})\left.\right|_{z=\pm a/2}=0,\label{eq24}
\end{equation}
where 
\begin{equation}
K^{(\pm)}=\gamma^0\left[{\rm e}^{{\rm i}\varphi_{\pm}\gamma^5}\cos\theta_{\pm} + (\gamma^1\cos\varsigma_{\pm} + 
\gamma^2\sin\varsigma_{\pm})\sin\theta_{\pm}\right]{\rm e}^{\pm{\rm i}\tilde{\varphi}_{\pm}\gamma^0\gamma^z}, \label{eq25}
\end{equation}
coordinates $\mathbf{r}=(x,\,y,\,z)$ are chosen in such a way that $x$ and $y$ are tangential to the boundary, while $z$ is 
normal to it, and the position of $\partial\Omega^{(\pm)}$ is identified with $z=\pm a/2$. The confinement of matter inside the 
slab means that the vector bilinear, $\chi^{\dag}(\mathbf{r})\gamma^0{\gamma}^z\chi(\mathbf{r})$, vanishes at the slab boundaries,
\begin{equation}
\chi^{\dag}(\mathbf{r})\gamma^0{\gamma}^z\chi(\mathbf{r})\left.\right|_{z=\pm a/2}=0,\label{eq26}
\end{equation}
and this is ensured by condition (24). As to the axial bilinear, 
$\chi^{\dag}(\mathbf{r})\gamma^0{\gamma}^z\gamma^5\chi(\mathbf{r})$, it vanishes at the slab boundaries,
\begin{equation}
\chi^{\dag}(\mathbf{r})\gamma^0{\gamma}^z\gamma^5\chi(\mathbf{r})\left.\right|_{z=\pm a/2}=0,\label{eq27}
\end{equation}
in the case of $\theta_+ = \theta_- = \pi/2$ only, that is due to relation 
\begin{equation}
[K^{(\pm)}\left.\right|_{\theta_{\pm} = \pi/2}, \gamma^5]_- = 0. \label{eq28}
\end{equation}
However, there is a symmetry with respect to rotations around a normal to the slab, and the cases differing by values of 
$\varsigma_+$ and $\varsigma_-$ are physically indistinguishable, since they are related by such a rotation. The only way to 
avoid the unphysical degeneracy of boundary conditions with different values of $\varsigma_+$ and $\varsigma_-$ is to fix 
$\theta_+=\theta_-=0$. Then $\chi^{\dag}(\mathbf{r})\gamma^0{\gamma}^z\gamma^5\chi(\mathbf{r})$ is nonvanishing at the slab 
boundaries, and the boundary condition takes form
\begin{equation}
\left\{I-\gamma^0\exp\left[{\rm i}\left(\varphi_\pm\gamma^5\pm\tilde{\varphi}_\pm\gamma^0\gamma^z\right)\right]\right\}
\chi(\mathbf{r})\left.\right|_{z=\pm a/2}=0. \label{eq29}
\end{equation}
Condition (29) determines the spectrum of the wave number vector in the $z$ direction, $k_l$. The requirement that this spectrum 
be real and unambiguous yields constraint (see \cite{Si2, Si3})
\begin{equation}
\varphi_+=\varphi_-=\varphi, \quad \tilde{\varphi}_+=\tilde{\varphi}_-=\tilde{\varphi};\label{eq30}
\end{equation}
then the $k_l$ spectrum is determined implicitly from relation 
\begin{equation}
k_l\sin\tilde{\varphi}\cos(k_l a)+(E_{... l}\cos\tilde{\varphi}-M\cos\varphi)\sin(k_l a)=0, \label{eq31}
\end{equation}
where $M$ is the mass of the spinor matter field and $E_{... l}$ is the energy of the one-particle state.
In the case of the massless spinor matter field, $M=0$, and the background uniform magnetic field perpendicular to the slab,
$E_{... l}$ takes the form of $E_{n l}$ (19), and relation (31) is reduced to
\begin{equation}
k_l\sin\tilde{\varphi}\cos(k_l a)+E_{n l}\cos\tilde{\varphi}\sin(k_l a)=0, \label{eq32}
\end{equation}
depending on one parameter only, although the boundary condition depends on two parameters,
\begin{equation}
\left\{I-\gamma^0\exp\left[{\rm i}\left(\varphi\gamma^5 \pm  \tilde{\varphi}\gamma^0\gamma^z\right)\right]\right\}
\chi(\mathbf{r})|_{z=\pm a/2}=0. \label{eq33}
\end{equation}

\section{Chiral effects} 

As was already mentioned, we are interested in the case of the uniform magnetic field directed perpendicular to the slab, 
${\bf B}=(0,0,B)$. The explicit form of the solution to Dirac equation (4) with Hamiltonian (18) in gauge ${\bf A}=(-yB,0,0)$ is 
given in Appendix A. It is straightforward to check the validity of relations
\begin{equation}
\left\langle j;qnk_l|{\bf r}\right\rangle\gamma^0\gamma^z\left\langle {\bf r}|j;qnk_l\right\rangle=
\left\langle j;qnk_l|{\bf r}\right\rangle\gamma^5\left\langle {\bf r}|j;qnk_l\right\rangle=0 \label{eq34}
\end{equation}
and
\begin{equation}
\int\limits_{-\infty}^{\infty}{\rm d}q\left\langle j;qnk_l|{\bf r}\right\rangle\gamma^0\gamma^x\left\langle {\bf r}|j;qnk_l
\right\rangle 
=\int\limits_{-\infty}^{\infty}{\rm d}q\left\langle j;qnk_l|{\bf r}\right\rangle\gamma^0\gamma^y\left\langle {\bf r}|j;qnk_l
\right\rangle=0,\label{eq35}
\end{equation}
which result in the vanishing of the vector current and axial charge densities,
\begin{equation}
{\bf J}=J^{05}=0. \label{eq36}
\end{equation}
Similarly, as a consequence of relation 
\begin{equation}
\int\limits_{-\infty}^{\infty}{\rm d}q\left\langle j;qnk_l|{\bf r}\right\rangle\gamma^0\gamma^x\gamma^5
\left\langle {\bf r}|j;qnk_l\right\rangle 
=\int\limits_{-\infty}^{\infty}{\rm d}q\left\langle j;qnk_l|{\bf r}\right\rangle\gamma^0\gamma^y\gamma^5
\left\langle {\bf r}|j;qnk_l\right\rangle=0,\label{eq37}
\end{equation}
the components of the axial current density, which are orthogonal to the direction of the magnetic field, vanish as well,
\begin{equation}
J^{x5}=J^{y5}=0.\label{eq38}
\end{equation}
As to the component of the axial current density, which is along the magnetic field, only the lowest Landau level ($n=0$) 
contributes to it, and that is due to relations
\begin{equation}
\int\limits_{-\infty}^{\infty}{\rm d}q\left\langle 0;q0k_l|{\bf r}\right\rangle\gamma^0\gamma^z\gamma^5
\left\langle {\bf r}|0;q0k_l\right\rangle =-\frac{eB}{2\pi a} \label{eq39}
\end{equation}
and
\begin{equation}
\sum\limits_{j=1,2}\int\limits_{-\infty}^{\infty}{\rm d}q\left\langle j;qnk_l|{\bf r}\right\rangle\gamma^0\gamma^z\gamma^5
\left\langle {\bf r}|j;qnk_l\right\rangle =0.\label{eq40}
\end{equation}
The spectrum of the wave number vector along the magnetic field is determined from (32) at $n=0$, i.e.,
\begin{equation}
k_{l}^{(\pm)}=(l\pi\mp\tilde{\varphi})/a, \quad l\in \mathbb{Z}, \quad k_l^{(\pm)}>0,\label{eq41}
\end{equation}
where the upper (lower) sign corresponds to $E_{0l}>0$ ($E_{0l}<0$) and $\mathbb{Z}$ is the set of integer numbers. Hence, the 
$z$ component of the axial current density is
\begin{equation}
J^{z5}=-\frac{eB}{2\pi a}\left[\sum\limits_{k_l^{(+)}>0}f_+(k_l^{(+)}) - \sum\limits_{k_l^{(-)}>0}f_-(k_l^{(-)}) -
\frac{1}{2}\sum\limits_{k_{l}^{(+)}>0}1 + \frac{1}{2}\sum\limits_{k_{l}^{(-)}>0}1\right],\label{eq42}
\end{equation}
where
\begin{equation}
f_\pm(k)=\left[e^{(k\mp\mu)/T}+1\right]^{-1}, \label{eq43}
\end{equation}
and the two last sums which are independent of temperature and chemical potential correspond to the contribution of the vacuum 
fluctuations.

The sums in (42) are transformed in Appendix B to render expression (B.21) which can be rewritten as
\begin{equation}
\sum\limits_{k_l^{(+)}>0}f_+(k_l^{(+)})-\sum\limits_{k_l^{(-)}>0}f_-(k_l^{(-)})=
{\rm sgn}(\mu)\left[F_1(s;t)+F_2(s;t)\right]\label{eq44}
\end{equation}
where
\begin{equation}
F_1(s;t)=\frac 12\sum\limits_{-s<s_m<s}{\rm tanh}\left[(s_m+s)(2t)^{-1}\right],\label{eq45}
\end{equation}
\begin{equation}
F_2(s;t)=\frac 12\sum\limits_{s_m>s}\left\{{\rm tanh}\left[(s_m+s)(2t)^{-1}\right]-
{\rm tanh}\left[(s_m-s)(2t)^{-1}\right]\right\},\label{eq46}
\end{equation}
and notations for dimensionless quantities are introduced as 
\begin{equation}
s=|\mu|a+{\rm sgn}(\mu)\left[\tilde{\varphi}-{\rm sgn}(\tilde{\varphi})\frac{\pi}{2}\right], \quad t=Ta, \quad 
s_m=\left(m+\frac{1}{2}\right)\pi,\label{eq47}
\end{equation}
$m \in \mathbb{Z}$. The finite sum in $F_1(s;t)$ and the infinite sum in $F_2(s;t)$ are calculated in Appendix B, yielding 
\begin{eqnarray}
F_1(s;t)=\frac{1}{\pi}{\rm sinh}\left[(2s+\pi)/(2t)\right]\int\limits_{0}^{\infty}{\rm d}v
\biggl(\left[\cos(2s)+e^{-2v}\right]{\rm sinh}\left[(2s-\pi)/(2t)\right]\sin(v/t)\biggr.\nonumber 
\\ \biggl. -\sin(2s)\left\{{\rm cosh}\left[(2s+\pi)/(2t)\right]+{\rm cosh}\left[(2s-\pi)/(2t)\right]\cos(v/t)\right\}\biggr)
\left[\cos(2s)+{\rm cosh}(2v)\right]^{-1}\nonumber \\
\times\left\{{\rm cosh}(2s/t){\rm cosh}(\pi/t)+2{\rm cosh}\left[(2s+\pi)/(2t)\right]{\rm cosh}\left[(2s-\pi)/(2t)\right]\cos(v/t)+\cos^2(v/t)\right\}^{-1}\nonumber \\
+\frac{t}{\pi}{\rm ln}\frac{{\rm cosh}(s/t)}{{\rm cosh}[\pi/(2t)]}+\frac{1}{2}{\rm tanh}\left\{\left[{\rm arctan}({\rm tan}s)
+\frac{\pi}{2}\right]/(2t)\right\}\label{eq48}
\end{eqnarray}
and
\begin{eqnarray}
F_2(s;t)=-\frac{1}{\pi}{\rm sinh}\left[(2s-\pi)/(2t)\right]\int\limits_{0}^{\infty}{\rm d}v
\biggl(\left[\cos(2s)+e^{-2v}\right]{\rm sinh}\left[(2s+\pi)/(2t)\right]\sin(v/t)\biggr.\nonumber 
\\ \biggl.-\sin(2s)\left\{{\rm cosh}\left[(2s-\pi)/(2t)\right]+{\rm cosh}\left[(2s+\pi)/(2t)\right]\cos(v/t)\right\}\biggr)
\left[\cos(2s)+{\rm cosh}(2v)\right]^{-1}\nonumber \\
\times\left\{{\rm cosh}(2s/t){\rm cosh}(\pi/t)+2{\rm cosh}\left[(2s+\pi)/(2t)\right]{\rm cosh}
\left[(2s-\pi)/(2t)\right]\cos(v/t)+\cos^2(v/t)\right\}^{-1}\nonumber \\
+\frac{s}{\pi}-\frac{t}{\pi}{\rm ln}\frac{{\rm cosh}(s/t)}{{\rm cosh}[\pi/(2t)]}
+\frac{1}{2}{\rm tanh}\left\{\left[{\rm arctan}({\rm tan}s)-\frac{\pi}{2}\right]/(2t)\right\}.\label{eq49}
\end{eqnarray}
The asymptotics of $F_1(s;t)$ and $F_2(s;t)$ at $t\rightarrow 0$ and $t\rightarrow \infty$ are given in Appendix B, see (B.24) and 
(B.27). Also, one can get
\begin{equation}
\lim\limits_{a\rightarrow \infty}\frac{1}{a}F_1(|\mu|a;Ta)=0\label{eq50}
\end{equation}
and
\begin{equation}
\lim\limits_{a\rightarrow \infty}\frac{1}{a}F_2(|\mu|a;Ta)=\frac{|\mu|}{\pi}.\label{eq51}
\end{equation}

Defining
\begin{equation}
F(s;t)=F_1 (s;t) + F_2 (s;t),\label{eq52}
\end{equation}
we obtain
\begin{eqnarray}
F(s;t)=\frac{s}{\pi}-\frac{1}{\pi}\int\limits_{0}^{\infty}{{\rm d}v\,\frac{\sin(2s){\rm sinh}(\pi/t)}
{[\cos(2s)+{\rm cosh}(2v)][{\rm cosh}(\pi/t)+\cos(v/t)]}}\nonumber \\ 
+\frac{{\rm sinh} \left\{[{\rm arctan} ({\rm tan} s)]/t\right\}}{{\rm cosh}[\pi/(2t)]+{\rm cosh}
\left\{[{\rm arctan} ({\rm tan} s)]/t\right\}}.\label{eq53}
\end{eqnarray}
Thus, the axial current density along the magnetic field is given by expression 
\begin{equation}
J^{z5}=-\frac{eB}{2\pi a}\Biggl\{{\rm sgn}(\mu)F\Biggl(|\mu|a + {\rm sgn}(\mu)\left[\tilde{\varphi}-
{\rm sgn}(\tilde{\varphi}){\pi}/{2}\right];Ta\Biggr) - \frac{1}{\pi}\left[\tilde{\varphi}-
{\rm sgn}(\tilde{\varphi}){\pi}/{2}\right]\Biggr\} , \label{eq54}
\end{equation}
where $F(s;t)$ is given by (53) and the contribution of the vacuum fluctuations,
\begin{equation}
-\frac{1}{2}\sum\limits_{k_{l}^{(+)}>0}1 + \frac{1}{2}\sum\limits_{k_{l}^{(-)}>0}1 = - \frac{\tilde{\varphi}}{\pi} + 
\frac{1}{2}{\rm sgn}(\tilde{\varphi}), \label{eq55}
\end{equation}
is taken into account.

In view of relations (50) and (51),
the case of a magnetic field filling the whole (infinite) space \cite{Vil,Son,Met} is obtained from (54) as a limiting 
case [cf. (1)],
\begin{equation}
\lim\limits_{a\rightarrow \infty}J^{z5}=-\frac{eB}{2\pi^2} \mu.\label{eq56}
\end{equation}
Unlike this unrealistic case, the realistic case of a magnetic field confined to a slab of finite width is temperature dependent, 
see (53) and (54). In particular, we get
\begin{equation}
\lim\limits_{T\rightarrow 0}J^{z5}=-\frac{eB}{2\pi a}\Biggl[{\rm sgn}(\mu)\left[\!\!\left[\frac{|\mu|a+
{\rm sgn}(\mu)\tilde{\varphi}}{\pi} 
+ \Theta(-\mu\tilde{\varphi})\right]\!\!\right] - \frac{\tilde{\varphi}}{\pi} + 
\frac{1}{2}{\rm sgn}(\tilde{\varphi})\Biggr] \label{eq57}
\end{equation}
and
\begin{equation}
\lim\limits_{T\rightarrow \infty}J^{z5}=-\frac{eB}{2\pi^2} \mu; \label{eq58}
\end{equation} 
here $[\![u]\!]$ denotes the integer part of quantity $u$ (i.e. the integer which is less than or equal 
to $u$), and $\Theta(u)=\frac{1}{2}[1+{\rm sgn}(u)]$ is the step function. As follows from (54), the boundary condition that is 
parametrized by $\tilde{\varphi}$ can serve as a source which is additional to the spinor matter density: the contribution of the 
boundary to the axial current effectively enhances or diminishes the contribution of the chemical potential. Because of the boundary 
condition, the chiral separation effect can be nonvanishing even at zero chemical potential,
\begin{equation}
J^{z5}|_{\mu=0}=-\frac{eB}{2\pi a}\Biggl\{F(\tilde{\varphi}-{\rm sgn}(\tilde{\varphi})\pi/2; Ta) - 
\frac{1}{\pi}\left[\tilde{\varphi}-{\rm sgn}(\tilde{\varphi}){\pi}/{2}\right]\Biggr\}; \label{eq59}
\end{equation}
the latter vanishes in the limit of infinite temperature, 
\begin{equation}
\lim\limits_{T \rightarrow \infty}J^{z5}|_{\mu=0}=0.\label{eq60}
\end{equation}
The trivial boundary condition, $\tilde{\varphi}=-\pi/2$, yields spectrum $k_l=(l+\frac 12)\frac{\pi}{a}$ 
$\quad$ ($l=0,1,2,\ldots$), which is the same in the setups of both bag models \cite{Bog} and \cite{Cho1}, and the axial current 
density at zero temperature for this case was obtained in \cite{Gor},
\begin{equation}
J^{z5}\left.\right|_{T=0, \,\, \tilde{\varphi}=-\pi/2} 
= - \frac{eB}{2\pi a}{\rm sgn}(\mu)\left[\!\!\left[ \frac{|\mu|a}{\pi}+\frac{1}{2} \right]\!\!\right]. \label{eq61}
\end{equation}
The "bosonic-type" spectrum, $k_l=l\frac{\pi}{a}$ $\quad$ ($l=0,1,2,\ldots$), is yielded by 
$\tilde{\varphi}=0$, and, due to the contribution of the vacuum fluctuations, the axial current density is continuous at 
this point,
\begin{equation}
\lim\limits_{\tilde{\varphi}\rightarrow 0_+}J^{z5} = \lim\limits_{\tilde{\varphi}\rightarrow 0_-}J^{z5}. \label{eq62}
\end{equation}
In a preliminary short letter communication announcing the results derived here (see \cite{Si4}), the contribution of the vacuum 
fluctuations was omitted. Because of this circumstance, the axial current density is discontinuous at $\tilde{\varphi}=0$ in 
\cite{Si4}. As well, the chiral separation effect at zero chemical potential disappears in \cite{Si4} at zero temperature, instead 
of (60). However, on theoretical grounds, the involvement of the vacuum fluctuations looks more consistent. Mention at least that, 
although the range of $\tilde{\varphi}$ is restricted to $-\frac{\pi}{2}\leq\tilde{\varphi}<\frac{\pi}{2}$ [see (22)], namely the 
account for the contribution of the vacuum fluctuations allows one to consider the axial current as a periodic continuous function 
of $\tilde{\varphi}$ on the whole unrestricted range, $-\infty<\tilde{\varphi}<\infty$.
 
Concluding this section, let us note that expression (1) is related to a quantum anomaly, being of topological nature which is 
revealed in the approach using the effective Lagrangian for Goldstone bosons of the spontaneously broken chiral symmetry 
\cite{Son}. Although the approach takes no account of temperature, it would be interesting to apply it to systems which are bounded 
in the direction of a magnetic field and to elucidate the role of the corresponding boundaries by determining the additional 
boundary terms (perhaps of the Chern-Simons type) to the effective Lagrangian. This will allow one to point at a complementary way for 
obtaining the chiral effects in a slab at zero temperature.

\section{Discussion and conclusion}

In the present paper, hot dense ultrarelativistic spinor matter in a slab under an influence of the transverse background magnetic 
field has been considered. An issue of boundary conditions plays the key role in this study. In the case of the quantized 
electromagnetic matter field, a choice of boundary conditions is motivated by material properties of bounding plates, 
and the conventional Casimir effect is different for different boundary conditions. For instance, it is attractive 
between the ideal-metal plates (i.e. made of material with an infinitely large magnitude of the dielectric permittivity), 
as well as between the plates made of material with an infinitely large magnitude of the magnetic 
permeability; meanwhile, it is repulsive between the ideal-metal plate and the infinitely permeable one; see, e.g., \cite{Bor}.
In the case of the quantized spinor matter field, nothing can be said about the ``material'' of boundaries, other than to admit 
that this material is impenetrable for spinor matter. Therefore, rather than attempting to model the microscopic details of 
the ``material'' of bounding plates, we have encoded the nature of boundaries in the values of parameters of the boundary condition ensuring 
the confinement of spinor matter inside the slab in the most general way. Mathematically acceptable, i.e. compatible with the 
self-adjointness of the Dirac Hamiltonian, is the eight-parametric boundary condition; see (24) and (25). The six-parametric 
boundary condition corresponding to $\theta_+ = \theta_- = \pi/2$ is consistent with the axial charge conservation; see (27). 
However, note that, as a massless spinor particle is reflected from an impenetrable boundary, its helicity is flipped. 
Since the chirality equals plus or minus the helicity, chiral symmetry has to be necessarily broken by the confining boundary 
condition. Thus the case of $\theta_+ = \theta_- = \pi/2$ is not acceptable on the physical grounds. Moreover, the requirement 
of an invariance with respect to rotations around a normal to the slab yields restriction $\theta_+ = \theta_- = 0$, and the 
boundary condition becomes a four-parametric one; see (29). A further physical requirement is that a standing wave inside the 
slab be unambiguously determined; this yields restriction (30), resulting finally in the two-parametric boundary condition [see 
(33)]. It should be noted that this condition breaks the time reversal symmetry, unless $\varphi = 0$. However, standing waves 
inside the slab are $\varphi$ independent [their wave number vectors are given by (41)], meaning that the physical effects 
of hot dense magnetized ultrarelativistic spinor matter in a slab are the same, irrespective 
of whether the time reversal symmetry is conserved or broken.

We have shown that the vector current and the axial charge are not induced by the background magnetic field in a slab; see (36). 
This is in contrast to the case of the unbounded medium, when such quantities are nonvanishing independently of 
chemical potential and temperature. Really, owing to chiral symmetry, 
\begin{equation}
 [H, \gamma^5]_- = 0,\label{eq63}
\end{equation}
one can introduce the following average [cf. (14)]:
\begin{equation}
\left\langle \hat{U}\right\rangle_{T,{\mu}_5}=-\frac{1}{2}{\rm tr}\left\langle {\bf r}|\Upsilon\tanh [(H-
{\mu}_5 \gamma^5)(2T)^{-1}]|{\bf r}\right\rangle, \label{eq64}
\end{equation}
where ${\mu}_5$ denotes the chiral chemical potential. Then a straightforward calculation with the use of wave functions of 
definite chiralities immediately yields
\begin{equation}
{\bf J}\equiv\left\langle \hat{U}\right\rangle_{T,{\mu}_5}\biggr|_{\Upsilon=\gamma^0\boldsymbol{\gamma}}=
-\frac{e{\bf B}}{2\pi^2}{\mu}_5   \label{eq65}
\end{equation}
and
\begin{eqnarray}
J^{05}\equiv\left\langle \hat{U}\right\rangle_{T,{\mu}_5}\biggr|_{\Upsilon=\gamma^5} \nonumber 
\\ =\frac{|eB|}{2\pi^2}\Biggl[{\mu}_5 +
2\sum\limits_{n=1}^{\infty}\int\limits_{0}^{\infty}{\rm d}k\,\frac{{\rm sinh}({\mu}_5/T)}{{\rm cosh}({\mu}_5/T)
+{\rm cosh}(\sqrt{k^2 + 2n|eB|}/T)}\Biggr];  \label{eq66}
\end{eqnarray}
note that only the lowest Landau level contributes to (65) which is known as the chiral magnetic effect \cite{Fuk}. 
In the case of a bounded medium, although relation (63) formally holds, one cannot determine a calculable version of average (64), 
because, as was already emphasized, the boundary condition breaks chiral symmetry. For instance, standing waves inside a slab are 
formed from counterpropagating waves of opposite chiralities. Thus, the chiral magnetic effect in a slab is prohibited by the 
confining boundary condition; see (26).

As to the chiral separation effect, we have shown that the axial current in a slab depends on both chemical potential and 
temperature; see (53) and (54). This is a main distinction from the unrealistic case of the unbounded medium, when the 
chiral separation effect is independent of temperature; see (1). At zero temperature, the chiral separation effect in a slab is 
characterized by a stepwise behaviour with the step width equal to $|eB|(2\pi a)^{-1}$; see (57). As the temperature increases, the 
steps are smoothed out, resulting in a linear behaviour at an infinitely large temperature; see (58). Another distinctive feature 
is the dependence of the chiral separation effect on the boundary parameter, $\tilde{\varphi}$: all values from range 
$-\frac{\pi}{2}<\tilde{\varphi}<\frac{\pi}{2}$ are allowable on an equal footing with value $\tilde{\varphi}=-\frac{\pi}{2}$. 
Only at an infinitely large temperature is the chiral separation effect independent of the boundary parameter, as well as of 
the slab width, coinciding with the chiral separation effect in the case of the unbounded medium. Otherwise, if the temperature is 
finite, a positive or negative term is added to the value of the chemical potential contributing to the axial current density, 
and the whole pattern is shifted along the line corresponding to the case of infinite temperature either to the right or to the 
left; see (54). In particular, the chiral separation effect can persist even at zero chemical potential: the axial 
current density is nonzero, being of the same (opposite) sign as $eB$ at $-\frac{\pi}{2}<\tilde{\varphi}<0$ 
($0<\tilde{\varphi}<\frac{\pi}{2}$); see (59). Perhaps, it is for the first time that such a mathematical entity as the 
self-adjoint extension parameter, $\tilde{\varphi}$, is to be determined experimentally (maybe, at least as an event-by-event 
fluctuation). It would be interesting to verify this in tabletop experiments with slabs of Dirac or Weyl semimetals in a 
transverse magnetic field.

\section*{Acknowledgments}

The work was supported by the National Academy of Sciences
of Ukraine (Project No.0112U000054), by the
Program of Fundamental Research of the Department of Physics and
Astronomy of the National Academy of Sciences of Ukraine (Project
No.0112U000056) and by the ICTP -- SEENET-MTP Grant PRJ-09
``Strings and Cosmology''.

\section*{Appendix A. Solution to the Dirac equation in a slab}

A solution to the Dirac equation in the background of a static uniform magnetic field is well described in the literature; see, 
e.g., \cite{Akhie}. In the case of a slab with a finite extent in the direction of the magnetic field, the general solution is 
a superposition of two plane waves counterpropagating along the magnetic field. Using the standard representation for the Dirac 
matrices,
$$
	\gamma^0=\begin{pmatrix}
I&0\\
0&-I
\end{pmatrix},\qquad
\boldsymbol{\gamma}=\begin{pmatrix}
0&\boldsymbol{\sigma}\\
-\boldsymbol{\sigma}&0
\end{pmatrix}
$$
($\sigma^1$, $\sigma^2$, and $\sigma^3$ are the Pauli matrices), choosing coordinate $z$ as directed along the magnetic field and 
taking $eB>0$ for definiteness, we get for the lowest Landau level ($n=0$):
$$
\left\langle {\bf r}|0;q0k_l\right\rangle|_{E_{0 l}>0}=\frac{e^{{\rm i}qx}}{2^{3/2}\pi}
\begin{pmatrix} 
C_0e^{{\rm i}k_lz}+\tilde{C}_0e^{-{\rm i}k_lz} \\ 0 \\
C_0e^{{\rm i}k_lz}-\tilde{C}_0e^{-{\rm i}k_lz} \\ 0
 \end{pmatrix}
\left(\frac{eB}{\pi}\right)^{1/4}\exp\left[-\frac{eB}{2}\left(y+\frac{q}{eB}\right)^2\right],\eqno(A.1)
$$
$$
\left\langle {\bf r}|0;q0k_l\right\rangle|_{E_{0 l}<0}=\frac{e^{-{\rm i}qx}}{2^{3/2}\pi}
\left(\begin{array}{c}
-C_0e^{{\rm i}k_lz}+\tilde{C}_0e^{-{\rm i}k_lz} \\ 0 \\
C_0e^{{\rm i}k_lz}+\tilde{C}_0e^{-{\rm i}k_lz} \\ 0
\end{array}\right)
\left(\frac{eB}{\pi}\right)^{1/4}\exp\left[-\frac{eB}{2}\left(y-\frac{q}{eB}\right)^2\right],\eqno{(A.2)}
$$
where coefficients $C_0$ and $\tilde{C}_0$ obey condition 
$$
|C_0|^2=|\tilde{C}_0|^2=\pi/a.\eqno{(A.3)}
$$
For the Landau levels with $n\geq 1$ we get two linearly independent solutions:
$$
\left\langle {\bf r}|1;qnk_l\right\rangle|_{E_{n l}>0}=\frac{e^{{\rm i}qx}}{2^{3/2}\pi}
\left(\begin{array}{c}
(C_1e^{{\rm i}k_lz}+\tilde{C}_1e^{-{\rm i}k_lz})Y_n^q(y) \\ 0 \\
\frac{k_l}{\omega_{nl}}(C_1e^{{\rm i}k_lz}-\tilde{C}_1e^{-{\rm i}k_lz})Y_n^a(y) \\
\frac{\sqrt{2neB}}{\omega_{nl}}(C_1e^{{\rm i}k_lz}+\tilde{C}_1e^{-{\rm i}k_lz})Y_{n-1}^q(y)
\end{array}\right),\eqno{(A.4)}
$$
$$
\left\langle {\bf r}|1;qnk_l\right\rangle|_{E_{n l}<0}=\frac{e^{-{\rm i}qx}}{2^{3/2}\pi}
\left(\begin{array}{c}
\frac{k_l}{\omega_{nl}}(-C_1e^{{\rm i}k_lz}+\tilde{C}_1e^{-{\rm i}k_lz})Y_n^{-q}(y) \\ 
\frac{\sqrt{2neB}}{\omega_{nl}}(-C_1e^{{\rm i}k_lz}-\tilde{C}_1e^{-{\rm i}k_lz})Y_{n-1}^{-q}(y) \\
(C_1e^{{\rm i}k_lz}+\tilde{C}_1e^{-{\rm i}k_lz})Y_{n}^{-q}(y) \\ 0
\end{array}\right),\eqno{(A.5)}
$$
and
$$
\left\langle {\bf r}|2;qnk_l\right\rangle|_{E_{n l}>0}=\frac{e^{{\rm i}qx}}{2^{3/2}\pi}
\left(\begin{array}{c}
0 \\ (C_2e^{{\rm i}k_lz}+\tilde{C}_2e^{-{\rm i}k_lz})Y_{n-1}^{q}(y) \\ 
\frac{\sqrt{2neB}}{\omega_{nl}}(C_2e^{{\rm i}k_lz}+\tilde{C}_2e^{-{\rm i}k_lz})Y_{n}^{q}(y) \\ \frac{k_l}{\omega_{nl}}(-C_2e^{{\rm i}k_lz}+\tilde{C}_2e^{-{\rm i}k_lz})Y_{n-1}^{q}(y)
\end{array}\right),\eqno{(A.6)}
$$
$$
\left\langle {\bf r}|2;qnk_l\right\rangle|_{E_{n l}<0}=\frac{e^{-{\rm i}qx}}{2^{3/2}\pi}
\left(\begin{array}{c}
\frac{\sqrt{2neB}}{\omega_{nl}}(-C_2e^{{\rm i}k_lz}-\tilde{C}_2e^{-{\rm i}k_lz})Y_{n}^{-q}(y) \\ \frac{k_l}{\omega_{nl}}(C_2e^{{\rm i}k_lz}-\tilde{C}_2e^{-{\rm i}k_lz})Y_{n-1}^{-q}(y) \\ 0 \\ 
(C_2e^{{\rm i}k_lz}+\tilde{C}_2e^{-{\rm i}k_lz})Y_{n-1}^{-q}(y)  \end{array}\right); \eqno{(A.7)}
$$
here
$$
Y_n^q(y)=\sqrt{\frac{(eB)^{1/2}}{2^nn!\pi^{1/2}}}\exp\left[-\frac{eB}{2}\left(y+\frac{q}{eB}\right)^2\right]H_n\left(\sqrt{eB}y+\frac{q}{\sqrt{eB}}\right),\eqno{(A.8)}
$$
$H_n(v)=(-1)^ne^{v^2}\frac{{\rm d}^n}{{\rm d}v^n}e^{-v^2}$ is the Hermite polynomial, function $Y_n^q(y)$ obeys orthonormalization 
condition 
$$
\int\limits_{-\infty}^{\infty}{\rm d}y\,Y_n^{q}(y)Y_{n'}^{q}(y)=\delta_{nn'},\eqno{(A.9)}
$$
and coefficients $C_j$ and $\tilde{C}_j$ $\quad$ ($j=1,2$) are chosen to obey condition 
$$
\left\{\begin{array}{c}
|C_j|^2=|\tilde{C}_j|^2=\pi/a, \\
C_j^*\tilde{C}_j + \tilde{C}_j^*C_j=0.
\end{array}\right.\eqno{(A.10)}
$$
The case of $eB<0$ is obtained by charge conjugation, i.e. changing $eB\rightarrow -eB$ and multiplying the complex conjugates of the above solutions by $\rm i\gamma^2$ (the energy sign is reversed).

Solutions with opposite signs of energy are orthogonal, 
$$
\int\limits_{\Omega}{\rm d}^3r\,\left\langle j;-qn-k_l|{\bf r}\right\rangle|_{E_{nl} \lessgtr 0}
\left\langle {\bf r}|j';q'n'k_{l'}\right\rangle|_{E_{nl} \gtrless 0} = 0, \quad j,j'=0,1,2,
\eqno{(A.11)}
$$
while solutions with the same sign of energy are orthogonal if an additional constraint is imposed,
$$
C_1^*\tilde{C}_2=\tilde{C}_1^*C_2.
\eqno{(A.12)}
$$
Thence, solutions (A.1), (A.2), (A.4)--(A.7) satisfy the requirements of orthonormality [cf. (5)],
$$
\int\limits_{\Omega}{\rm d}^3r\left\langle j;qnk_l|{\bf r}\right\rangle
\left\langle {\bf r}|j';q'n'k_{l'}\right\rangle=
\delta_{jj'}\delta_{nn'}\delta_{ll'}\delta(q-q'), \quad j,j'=0,1,2
\eqno{(A.13)}
$$
and completeness [cf. (6)],
$$
\int\limits_{-\infty}^{\infty}{\rm d}q\,\sum\limits_{l}\left(\left\langle {\bf r}|0;q0k_l\right\rangle
\left\langle 0;q0k_l|{\bf r}'\right\rangle+\sum\limits_{n=1}^{\infty}\sum\limits_{j=1,2}\left\langle {\bf r}|j;qnk_l\right\rangle\left\langle j;qnk_l|{\bf r}'\right\rangle\right)=I\delta({\bf r}-{\bf r}').
\eqno{(A.14)}
$$

\section*{Appendix B. Summation over the spectrum of the wave number vector of standing waves}

Condition (32) in the case of the lowest Landau level ($n=0$) is rewritten as 
$$
P_+(k_l^{(+)})=0, \quad E_{0l}>0.
\eqno{(B.1)}
$$
or
$$
P_-(k_l^{(-)})=0, \quad E_{0l}<0,
\eqno{(B.2)}
$$
where
$$
P_{\pm}(\omega)=\cos(\omega a)\pm \cot\tilde{\varphi}\sin(\omega a).
\eqno{(B.3)}
$$
The spectrum of the wave number vector in the direction of the magnetic field is determined by (B.1) or (B.2) and is given by 
(41). Using the Cauchy residue theorem, we get 
$$
\sum\limits_{k_l^{(+)}>0}f_+(k_l^{(+)})-\sum\limits_{k_l^{(-)}>0}f_-(k_l^{(-)})=\frac{a}{2\pi}\int\limits_{C_\subset}{\rm d}\omega\left[f_+(\omega)G_{\tilde{\varphi}}^{(+)}(\omega)-f_-(\omega)G_{\tilde{\varphi}}^{(-)}(\omega)\right],
\eqno{(B.4)}
$$
\begin{figure}
\includegraphics[width=390pt]{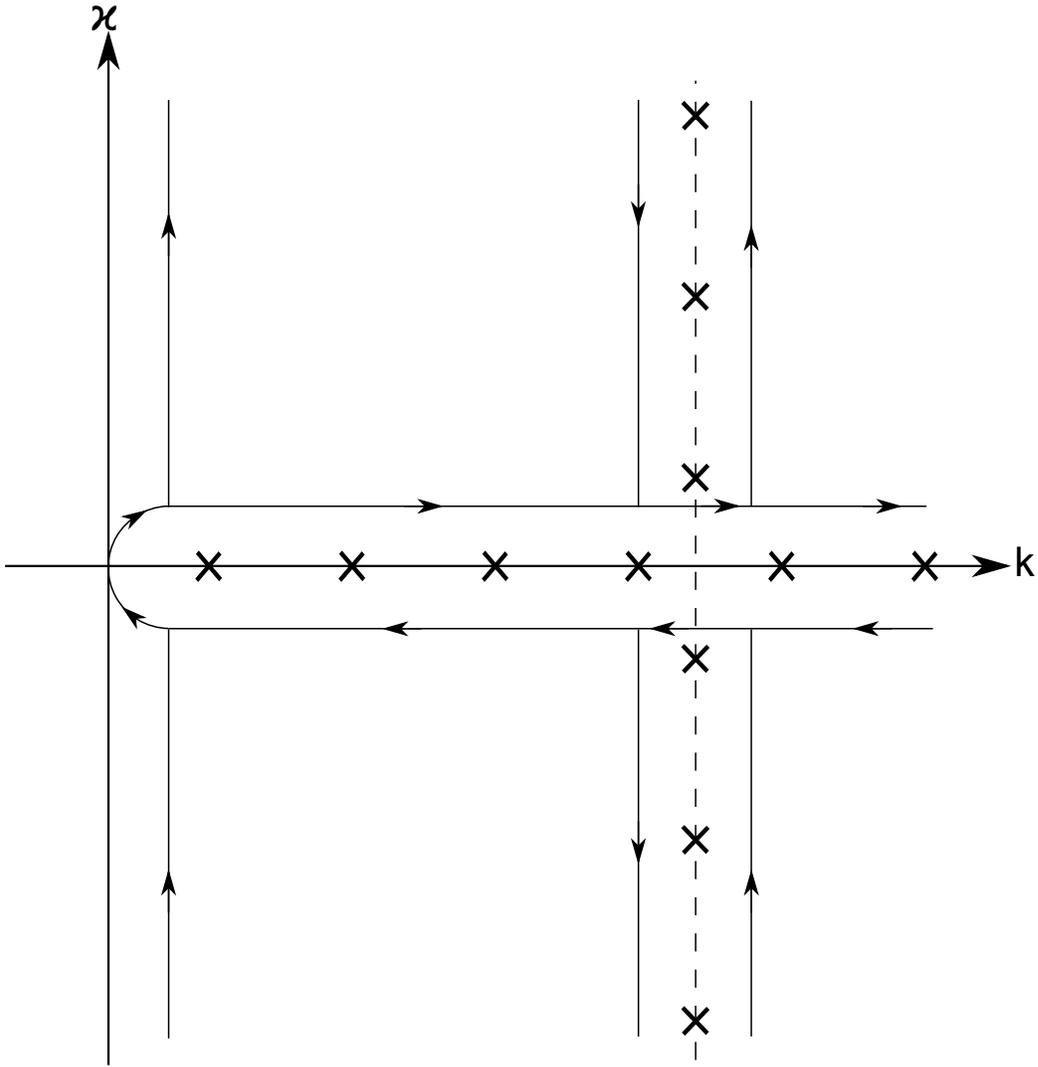}\\
\caption{Contour $C_\subset$ enclosing the positive real semiaxis can be continuously deformed into a contour consisting of 
vertical lines on the complex $\omega$ plane; positions of simple poles of the integrand are indicated by 
crosses.}\label{1}
\end{figure}\normalsize
where
$$
G_{\tilde{\varphi}}^{(\pm)}(\omega)=1+\frac{\rm i}{a}\,\frac{{\rm d}}{{\rm d}\omega}\ln P_\pm(\omega)
\eqno{(B.5)}
$$
and contour $C_\subset$ on the complex $\omega$ plane (${\rm Re}\,\omega=k$, ${\rm Im}\,\omega=\kappa$) encloses the positive 
real semiaxis; see Fig. 1. In view of relation
$$
G_{\tilde{\varphi}}^{(\pm)}(\omega)+G_{-\tilde{\varphi}}^{(\pm)}(-\omega)=2,
\eqno{(B.6)}
$$
we obtain
$$
\sum\limits_{k_l^{(+)}> 0}f_+(k_l^{(+)})-\sum\limits_{k_l^{(-)} > 0}
f_-(k_l^{(-)})=\frac{a}{\pi}\int\limits_{0}^{\infty}{\rm d}k[f_+(k)-f_-(k)]$$
$$
-\frac{a}{2\pi}\int\limits_{0}^{\infty}{\rm d}k\left[f_+(k)G_{-\tilde{\varphi}}^{(+)}(-k-{\rm i}\epsilon)-f_-(k)G_{-\tilde{\varphi}}^{(-)}(-k-{\rm i}\epsilon)\right. $$
$$ \left.+f_+(k)G_{\tilde{\varphi}}^{(+)}(k-{\rm i}\epsilon)-f_-(k)G_{\tilde{\varphi}}^{(-)}(k-{\rm i}\epsilon)\right],
 \eqno{(B.7)}
$$
where $\epsilon$ is positive real and infinitesimally small. The first term on the right-hand side of (B.7) is
$$
\frac{a}{\pi}\int\limits_{0}^{\infty}{\rm d}k\,\frac{{\rm sinh}(\mu/T)}{{\rm cosh}(\mu/T)+{\rm cosh}(k/T)}=\frac{\mu a}{\pi}.
\eqno{(B.8)}
$$
The integral in the second term on the right-hand side of (B.7) is transformed by continuous deformation of contour $C_\subset$ 
into a contour consisting of vertical lines; see Fig. 1 (the contribution of integrals over segments of a semicircle of infinite 
radius at ${\rm Re}\,\omega>0$ is exponentially damped). The contribution of the integral over the line which is infinitesimally 
close to ${\rm Re}\,\omega =0$ is 
$$
-\frac{{\rm i}a}{2\pi}\int\limits_{0}^{\infty}{\rm d}\kappa \left[f_+(\epsilon+{\rm i}\kappa)G_{-\tilde{\varphi}}^{(+)}(-{\rm i}\kappa)-f_-(\epsilon+{\rm i}\kappa)G_{-\tilde{\varphi}}^{(-)}(-{\rm i}\kappa)\right.$$
$$ 
\left.-f_+(\epsilon-{\rm i}\kappa)G_{\tilde{\varphi}}^{(+)}(-{\rm i}\kappa)+f_-(\epsilon-{\rm i}\kappa)G_{\tilde{\varphi}}^{(-)}(-{\rm i}\kappa)\right]$$
$$=-\frac{a}{\pi}\int\limits_{0}^{\infty}{\rm d}\kappa\,\frac{\sin(2\tilde{\varphi})}{-\cos(2\tilde{\varphi})+{\rm cosh}(2\kappa a)}=\frac{1}{\pi}\left[\tilde{\varphi}-{\rm sgn}(\tilde{\varphi})\frac{\pi}{2}\right],
 \eqno{(B.9)}
$$
where the use is made of relation
$$
\int\limits_{0}^{\infty}{\rm d}\eta\,\frac{\sin x}{\cos x+{\rm cosh}\eta}=2\arctan\left(\tan\frac{x}{2}\right).\eqno{(B.10)}
$$
The contribution of the integral over the lines which are infinitesimally close to ${\rm Re}\,\omega =|\mu|$ is reduced to the 
contribution of simple poles at $\kappa=\pm(2l'+1)\pi T$ $\quad$ ($l'=0,1,2,\ldots$),
\newpage
$$
-\frac{{\rm i}a}{2\pi}{\rm sgn}(\mu)\int\limits_{0}^{\infty}{\rm d}\kappa 
\left\{\left[f_{{\rm sgn}(\mu)}(|\mu|+\epsilon+{\rm i}\kappa)-f_{{\rm sgn}(\mu)}(|\mu|-\epsilon+{\rm i}\kappa)\right]
G_{-\tilde{\varphi}}^{({\rm sgn}(\mu))}(-|\mu|-{\rm i}\kappa)\right.$$
$$ 
\left. - \left[f_{{\rm sgn}(\mu)}(|\mu|+\epsilon-{\rm i}\kappa)-f_{{\rm sgn}(\mu)}(|\mu|-\epsilon-{\rm i}\kappa)\right]
G_{\tilde{\varphi}}^{({\rm sgn}(\mu))}(|\mu|-{\rm i}\kappa)\right\}$$
$$
={\rm sgn}(\mu)\sum\limits_{k_{l'}>0}\tilde{f}(k_{l'}), \quad k_{l'}=\left(l'+\frac{1}{2}\right)\frac{\pi}{a}, \quad 
l'=0,1,2,\ldots, \eqno{(B.11)}
$$
where
$$
\tilde{f}(\omega)=-\frac{2Ta\, \sin\{2|\mu|a+{\rm sgn}(\mu)[2\tilde{\varphi}-{\rm sgn}(\tilde{\varphi})\pi]\}}
{\cos\{2|\mu|a+{\rm sgn}(\mu)[2\tilde{\varphi}-{\rm sgn}(\tilde{\varphi})\pi]\}+{\rm cosh}(4\omega Ta^2)}. \eqno{(B.12)}
$$
\begin{figure}
\includegraphics[width=390pt]{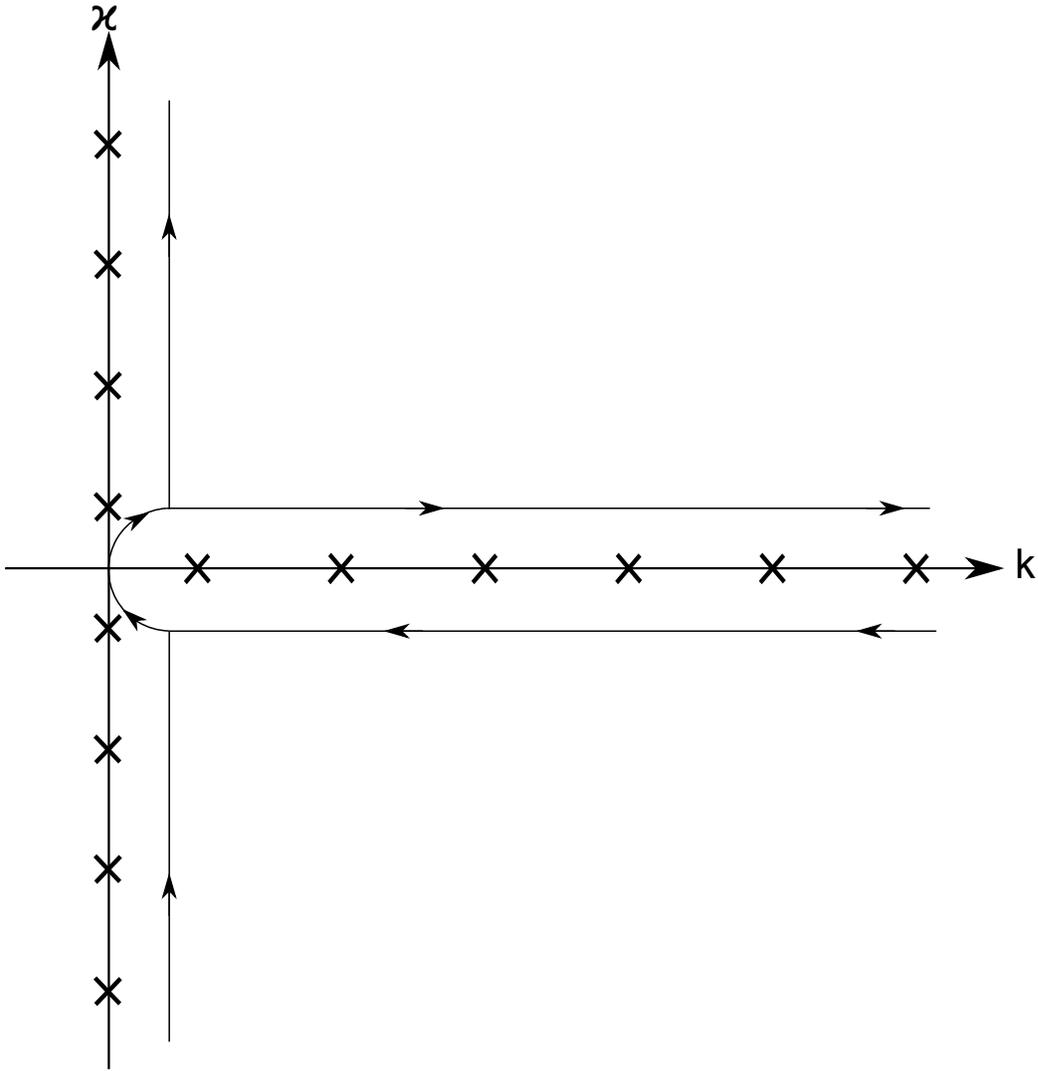}\\
\caption{Poles on the complex plane are on the imaginary axis. Contour enclosing infinite number of poles on the positive real 
semiaxis is deformed into a contour which is infinitesimally close from the right to the imaginary axis.}\label{2}
\end{figure}\normalsize
Then the sum on the right-hand side of (B.11) is presented as an integral of $2\tilde{f}(\omega)(e^{2{\rm i}\omega a}+1)^{-1}$ 
over a contour enclosing the positive real semiaxis on the complex $\omega$ plane; see Fig. 2. With the use of relation 
$$
(e^{2{\rm i}\omega a}+1)^{-1}+(e^{-2{\rm i}\omega a}+1)^{-1}=1,\eqno{(B.13)}
$$
we get
$$
\sum\limits_{k_{l'}>0}\tilde{f}(k_{l'})=\frac{a}{\pi}\int\limits_{0}^{\infty}{\rm d}k\,\tilde{f}(k)
-\frac{a}{\pi}\int\limits_{0}^{\infty}{\rm d}k\,\tilde{f}(k)\left\{\left[e^{-2{\rm i}(k+{\rm i}\epsilon)a}+1\right]^{-1}+\left[e^{2{\rm i}(k-{\rm i}\epsilon)a}+1\right]^{-1}\right\}. \eqno{(B.14)}
$$
Taking the first integral in (B.14) and transforming the second integral in (B.14) by continuous deformation on the complex 
$\omega$ plane into a contour which is infinitesimally close to the imaginary axis (see Fig. 2), we get 
$$
\sum\limits_{k_{l'}>0}\tilde{f}(k_{l'})=-\frac{1}{\pi}\arctan\left(\tan\left\{|\mu|a+{\rm sgn}(\mu)\left[\tilde{\varphi}
-{\rm sgn}(\tilde{\varphi})\frac{\pi}{2}\right]\right\}\right)$$
$$
-\frac{{\rm i}a}{\pi}\int\limits_{0}^{\infty}{\rm d}\kappa\left[\tilde{f}(\epsilon+{\rm i}\kappa)-\tilde{f}(\epsilon-{\rm i}\kappa)\right](e^{2\kappa a}+1)^{-1}. \eqno{(B.15)}
$$
Only simple poles (with half residues) contribute to the integral in (B.15), and thus we get 
\newpage
$$
-\frac{{\rm i}a}{\pi}\int\limits_{0}^{\infty}{\rm d}\kappa[\tilde{f}(\epsilon+{\rm i}\kappa)-\tilde{f}(\epsilon-{\rm i}\kappa)](e^{2\kappa a}+1)^{-1}$$
$$
=-\frac{1}{2}\sum\limits_{k_{m}^{(+)}>0}[1-{\rm tanh}(k_m^{(+)}a)]
+\frac{1}{2}\sum\limits_{k_{m}^{(-)}>0}[1-{\rm tanh}(k_m^{(-)}a)],
\eqno{(B.16)}
$$
where
$$
k_m^{(\pm)}=\left\{\left(m+\frac{1}{2}\right)\pi\pm\left[|\mu|a+{\rm sgn}(\mu)\tilde{\varphi}-
{\rm sgn}(\mu\tilde{\varphi})\frac{\pi}{2}\right]\right\}(2Ta^2)^{-1}, \quad  m \in \mathbb{Z}. \eqno{(B.17)}
$$
Accounting for (B.7)-(B.9), (B.11), (B.15) and (B.16), we obtain
$$
\sum\limits_{k_{l}^{(+)}>0}f_+(k_l^{(+)})-\sum\limits_{k_{l}^{(-)}>0}f_-(k_l^{(-)})={\rm sgn}(\mu)\left[\!\!\left[\frac{|\mu|a+
{\rm sgn}(\mu)\varphi}{\pi}+\Theta(-\mu\varphi)\right]\!\!\right]$$
$$
-\frac{1}{2}{\rm sgn}(\mu)\sum\limits_{k_{m}^{(+)}>0}\left[1-{\rm tanh}\left(k_m^{(+)}a\right)\right]+
\frac{1}{2}{\rm sgn}(\mu)\sum\limits_{k_{m}^{(-)}>0}\left[1-{\rm tanh}\left(k_m^{(-)}a\right)\right],\eqno{(B.18)}
$$
where the use is made of relation
$$
\frac{1}{\pi}\left[\mu a+\tilde{\varphi}-{\rm sgn}(\tilde{\varphi})\frac{\pi}{2}\right]
-\frac{1}{\pi}\arctan \left\{\tan\left[\mu a+\tilde{\varphi}-{\rm sgn}(\tilde{\varphi})\frac{\pi}{2}\right]\right\}$$
$$
={\rm sgn}(\mu)\left[\!\!\left[\frac{|\mu|a+{\rm sgn}(\mu)\tilde{\varphi}}{\pi}+\Theta(-\mu\tilde{\varphi})\right]\!\!\right]. 
\eqno{(B.19)}
$$
In view of relation
$$
\frac{1}{2}\sum\limits_{k_{m}^{(+)}>0}1-\frac{1}{2}\sum\limits_{k_{m}^{(-)}>0}1 =
\left[\!\!\left[\frac{|\mu|a+{\rm sgn}(\mu)\tilde{\varphi}}{\pi}+\Theta(-\mu\tilde{\varphi})\right]\!\!\right],\eqno{(B.20)}
$$
we get
$$
\sum\limits_{k_{l}^{(+)}>0}f_+(k_l^{(+)})-\sum\limits_{k_{l}^{(-)}>0}f_-(k_l^{(-)})$$
$$=\frac{1}{2}{\rm sgn}(\mu)\left[\sum\limits_{k_{m}^{(+)}>0}{\rm tanh}(k_m^{(+)}a)
-\sum\limits_{k_{m}^{(-)}>0}{\rm tanh}(k_m^{(-)}a)\right]; \eqno{(B.21)}
$$
note that each of the sums in (B.20) and on the right-hand side of (B.21) is divergent at large values of $m$, but their 
difference is finite. The latter relation is rewritten as (44) with (45)-(47).

Turning now to the analysis of $F_1(s,t)$ (45) and $F_2(s,t)$ (46), let us consider the case of $\frac{\pi}{2}<s<\infty$ first. 
Similar to the above, the sums in (45) and (46) are presented as integrals over contours enclosing simple poles on the complex 
plane. In the case of (45), the leftmost pole is separated, while all other poles are encircled by a conjoint contour; see 
Fig. 3. In this way we get
$$
F_1(s;t)=\frac{1}{2}\tanh\left\{\left[\arctan(\tan s)+\frac{\pi}{2}\right](2t)^{-1}\right\}
+\frac{1}{2\pi}\int\limits_{-s+\pi}^{s}{\rm d}w\,\tanh[(w+s)(2t)^{-1}]$$
$$
-\frac{1}{2\pi}\int\limits_{-s+\pi}^{s}{\rm d}w\,\tanh[(w+s)(2t)^{-1}]\left\{\left[e^{-2{\rm i}(w+{\rm i}\epsilon)}+1\right]^{-1}+\left[e^{2{\rm i}(w-{\rm i}\epsilon)}+1\right]^{-1}\right\}. \eqno{(B.22)}
$$
Taking the first integral and deforming the contour of integration in the second one into two vertical lines on the 
complex ($w + {\rm i}v$) plane (see Fig. 3), we get
$$
F_1(s;t)=\frac{1}{2}\tanh\left\{\left[\arctan(\tan s)+\frac{\pi}{2}\right](2t)^{-1}\right\}
+\frac{t}{\pi}\ln \frac{\cosh(st^{-1})}{\cosh[\pi(2t)^{-1}]}$$
$$
+\frac{1}{\pi}\int\limits_{0}^{\infty}{\rm d}v\,\left\{{\rm Im}\,\frac{\tan[(\pi+{\rm i}v)(2t)^{-1}]}
{e^{2v+2{\rm i}s}+1}-{\rm Im}\,\frac{\tan[(2s+{\rm i}v)(2t)^{-1}]}{e^{2v-2{\rm i}s}+1}\right\}. \eqno{(B.23)}
$$
As a result, formula (48) is obtained; we list here its asymptotics:
$$
\lim\limits_{t\rightarrow 0}F_1(s;\,t)=\left[\!\!\left[\frac{s}{\pi}+\frac{1}{2}\right]\!\!\right], \quad  
\lim\limits_{t\rightarrow \infty}F_1(s;\,t)=0. \eqno{(B.24)}
$$
\begin{figure}
\includegraphics[width=390pt]{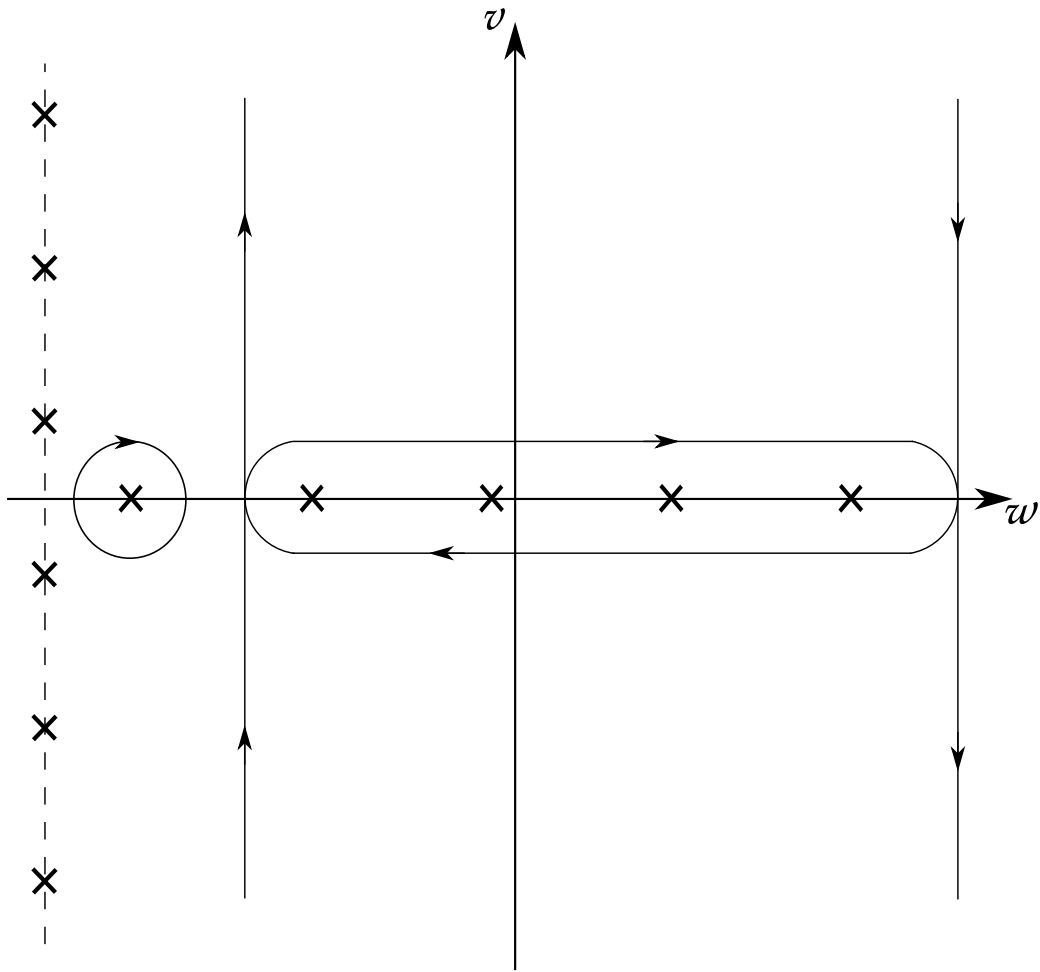}\\
\caption{Finite number of poles on the real axis at $-s < w < s$ is enclosed by two separate contours. Poles on the complex plane 
are on a vertical axis at $w = - s$. The right closed contour is deformed into a contour consisting of two vertical lines at 
$w = - s + \pi$ and  $w = s$.}\label{3}
\end{figure}\normalsize
\begin{figure}
\includegraphics[width=390pt]{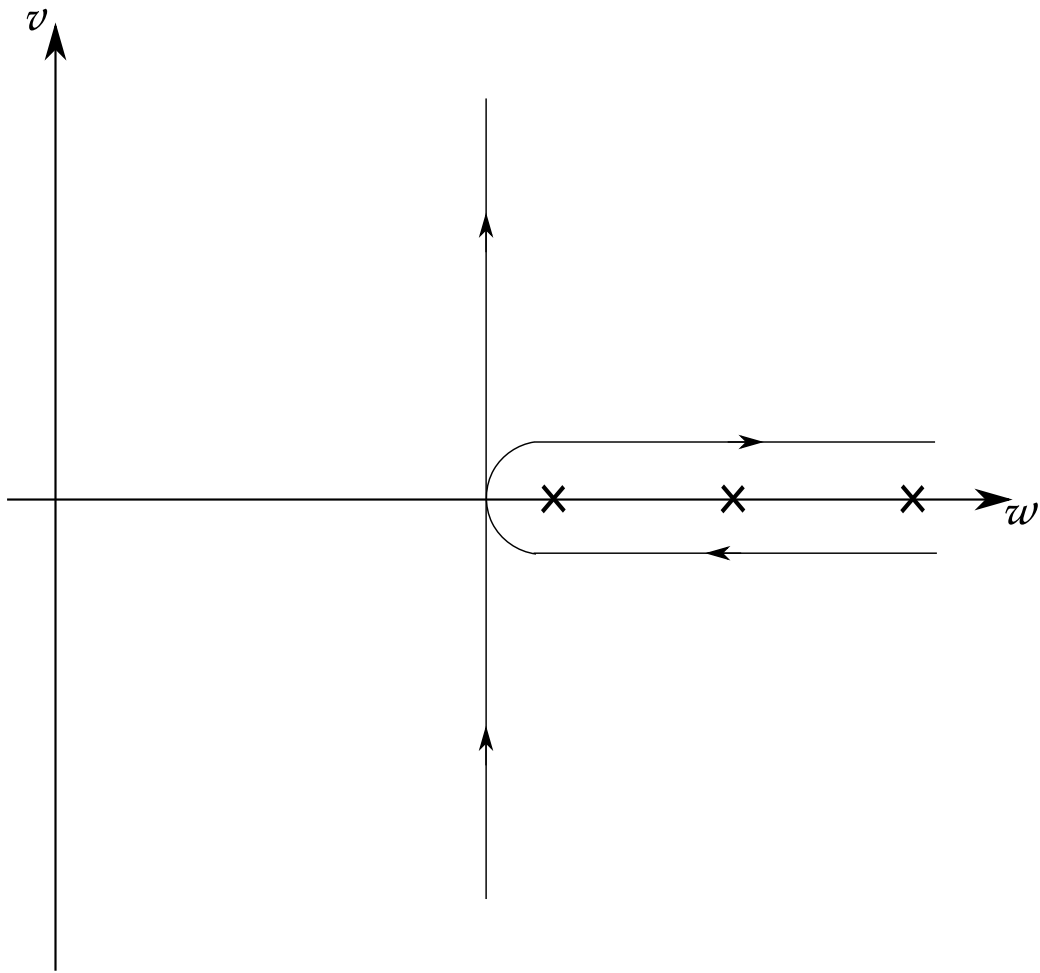}\\
\caption{Contour enclosing infinite number of poles on the real axis at $\quad$ $\quad$  $s < w < \infty$ is deformed into a contour consisting 
of a vertical line at $w = s$.}\label{4}
\end{figure}\normalsize
\begin{figure}
\includegraphics[width=390pt]{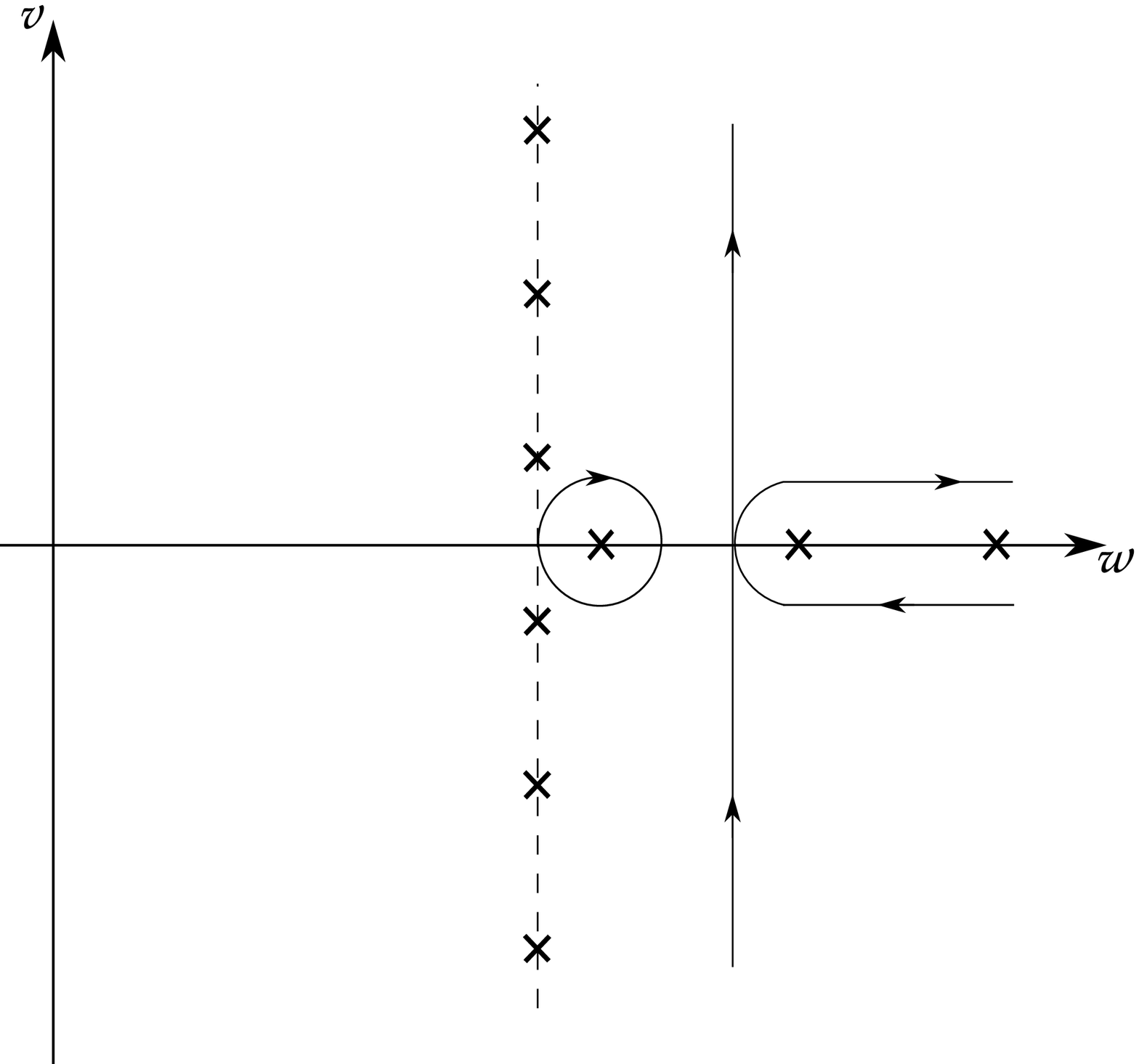}\\
\caption{Infinite number of poles on the real axis at $s < w < \infty$ is enclosed by two separate contours, the right one 
is deformed into a contour consisting of a vertical line at $w = s + \pi$. Poles on the complex plane are on a vertical 
line at $w = s$.}\label{5}
\end{figure}\normalsize

In the case of (46), the contour corresponding to the series with a positive sign encloses all poles (see Fig. 4), whereas the 
contour corresponding to the series with a negative sign is disjoint, separating the leftmost pole (see Fig. 5). In this way we get
\newpage
$$
F_2(s;t)=-\frac{1}{2}\tanh\left\{\left[\frac{\pi}{2}-\arctan(\tan s)\right](2t)^{-1}\right\}$$
$$
+\frac{1}{2\pi}\int\limits_{s}^{\infty}{\rm d}w\,\tanh[(w+s)(2t)^{-1}]-\frac{1}{2\pi}\int\limits_{s+\pi}^{\infty}{\rm d}w\,\tanh[(w-s)(2t)^{-1}]$$
$$
-\frac{1}{2\pi}\int\limits_{s}^{\infty}{\rm d}w\,\tanh[(w+s)(2t)^{-1}]\left\{\left[e^{-2{\rm i}(w+{\rm i}\epsilon)}+1\right]^{-1}+\left[e^{2{\rm i}(w-{\rm i}\epsilon)}+1\right]^{-1}\right\}$$
$$
+\frac{1}{2\pi}\int\limits_{s+\pi}^{\infty}{\rm d}w\,\tanh\left[(w-s)(2t)^{-1}\right]\left\{\left[e^{-2{\rm i}(w+{\rm i}\epsilon)}
+1\right]^{-1}+\left[e^{2{\rm i}(w-{\rm i}\epsilon)}+1\right]^{-1}\right\}. \eqno{(B.25)}
$$
The last two integrals are transformed by deforming the contour of integration into a vertical line on the complex 
($w + {\rm i}v$) plane (see Figs. 4 and 5), yielding
$$
F_2(s;t)=-\frac{1}{2}\tanh\left\{\left[\frac{\pi}{2}-\arctan(\tan s)\right](2t)^{-1}\right\}+\frac{s}{\pi}-\frac{t}{\pi}\ln\frac{\cosh(st^{-1})}{\cosh[\pi(2t)^{-1}]}$$
$$ + \frac{1}{\pi}
\int\limits_{0}^{\infty}{\rm d}v\,{\rm Im}
\left\{\tanh[(2s+{\rm i}v)(2t)^{-1}]-\tanh[(\pi+{\rm i}v)(2t)^{-1}]\right\}\left(e^{2v-2{\rm i}s}+1\right)^{-1}. \eqno{(B.26)}
$$
As a result, formula (49) is obtained; we list here its asymptotics:
$$
\lim\limits_{t\rightarrow 0}F_2(s;\,t)=0, \quad \lim\limits_{t\rightarrow \infty}F_2(s;\,t)=\frac{s}{\pi}. \eqno{(B.27)}
$$

Considering the case of $-\frac{\pi}{2}<s<\frac{\pi}{2}$, we note that $F_1(s;t)=0$ and 
$$
F_2(s;t)=\frac{1}{2}\sum\limits_{s_m>0}\left\{\tanh[(s_m+s)(2t)^{-1}]-\tanh[(s_m-s)(2t)^{-1}]\right\} \eqno{(B.28)}
$$
in this case. The contour of integration on the complex plane is chosen to be disjoint, separating the leftmost pole. As a 
result, formula (53) is obtained; this formula is the sum of $F_1(s;t)$ and $F_2(s;t)$ for the case of $\frac{\pi}{2}<s<\infty$. 
Note also, in view of (B.10), an alternative to (53) representation of $F(s;t)$, which is more relevant for the case of small $t$, 
\newpage
$$
F(s;t)=\left[\!\!\left[\frac{s}{\pi}+\frac{1}{2}\right]\!\!\right]+\frac{1}{\pi}\int\limits_{0}^{\infty}{\rm d}v\,
\frac{\sin(2s)[e^{-\pi/t}+\cos(v/t)]}{[\cos(2s)+\cosh(2v)][\cosh(\pi/t)+\cos(v/t)]}$$
$$
+\frac{\sinh\{[\arctan({\rm tan}s)]/t\}}{\cosh[\pi/(2t)]+\cosh\{[\arctan({\rm tan}s)]/t\}}. \eqno{(B.29)}
$$

Actually, in this appendix we have proven relation
$$
\sum\limits_{n \in \mathbb{Z}}\frac{y\sin x}{\cos x+\cosh[(2n+1)\pi y]}=\frac{1}{\pi} \int\limits_{0}^{\infty}{\rm d}\eta \,\frac{\sin x\sinh(2\pi/y)}{(\cos x+\cosh \eta)
[\cosh(2\pi/y)+\cos(\eta/y)]}$$
$$
-\frac{2\sinh\{2[\arctan\left(\tan\frac{x}{2}\right)]/y\}}{\cosh(\pi/y)+\cosh\{2[\arctan\left(\tan\frac{x}{2}\right)]/y\}}, \eqno{(B.30)}
$$
which may have diverse applications in fermion field theory at finite temperature and chemical potential.

\end{document}